\documentclass[11pt,a4paper]{article}
\pdfoutput=1
\usepackage{jcappub}
\usepackage{latexsym}
\usepackage{amssymb}
\usepackage{amsmath}
\usepackage{epsfig}
\usepackage{multirow}

\newcommand{\grad}{\ensuremath{\vec{\nabla}}}

\newcommand{\del}{\ensuremath{\partial}}
\renewcommand{\O}{{\cal O}}

\title{The Parametrized Post-Newtonian-Vainshteinian Formalism}

\author[a]{A. Avilez-Lopez,}
\author[a]{A. Padilla,}
\author[a]{Paul M. Saffin,}
\author[a,b]{C. Skordis.}
\affiliation[a]{School of Physics and Astronomy, University of Nottingham, University Park, Nottingham NG7 2RD, United Kingdom}
\affiliation[b]{Department of Physics, University of Cyprus, 1 University Avenue, Nicosia 2109, Cyprus}
\emailAdd{ppxaaa@nottingham.ac.uk}
\emailAdd{antonio.padilla@nottingham.ac.uk}
\emailAdd{paul.saffin@nottingham.ac.uk}
\emailAdd{skordis@ucy.ac.cy}

\abstract{
Light degrees of freedom that modify gravity on cosmological scales must be ``screened" on solar system scales in order to be compatible with data. The Vainshtein mechanism  achieves this through a breakdown of classical perturbation theory, as  large interactions involving new degrees of freedom become important below the so-called Vainshtein radius.
We begin to develop an extension of the Parameterized Post-Newtonian (PPN) formalism that is able to handle Vainshteinian corrections. We argue that 
theories with a unique Vainshtein scale must be expanded using two small parameters.  In this Parameterized Post-Newtonian-Vainshteinian (PPNV) expansion, 
the primary expansion parameter that controls the PPN order is, as usual, the velocity $v$. The secondary expansion parameter, $\alpha$, controls the strength of the 
Vainshteinian correction and is a theory-specific combination of the Schwarzschild radius and the Vainshtein radius of the source that is independent of its mass.  
We present the general framework and apply it to Cubic Galileon theory  both inside and outside the Vainshtein radius. 
The PPNV framework can be used to determine the compatibility of such theories with solar system and other strong-field data.
}


\begin{document}

\maketitle

\section{Introduction}
 General Relativity (GR) is a very successful theory. Since its inception, it has not only led to the prediction of new physical phenomena,
 but has also  successfully passed all experimental tests to this day~\cite{Will2014a,Will2014b}. But the question of whether GR is the correct theory
for describing all gravitational phenomena in Nature still stands. Indeed, experimental gravitation is a very active field in physics, aimed 
at increasing the precision of gravitational tests, but also at pushing the boundaries where these tests have been carried out.

In order to test gravity, one may simply take a gravitational theory, calculate its predictions for the tests of interest, and compare with observations. However,
as the ``space'' of  gravitational theories is vast, perhaps infinite, this is not a very economical process. A better way is to construct a ``master theory'', a framework,
that encompasses a wide range of theories.  Different theories would then correspond to a specific set of parameters of this framework.
Examples are the Parametrized Post-Newtonian (PPN)~\cite{Nordtvedt1968,Will1971,WillNordtvedt1972a}, the Parametrized Post-Keplerian (PPK)~\cite{DamourTaylor1992}, 
and the Parametrized Post-Einsteinian (PPE)~\cite{YunesPretorius2009,CornishEtAl2011,LoutrelYunesPretorius2014} formalisms.
In the cosmological regime we also have the various simplified parametrized approaches~\cite{ZhangEtAl2007,SilvestriPogosianBuniy2013}, 
 the small-scale parametrized approach~\cite{AminWagonerBlandford2007}
as well as more complete frameworks such as the Hu-Sawicki Parametrized Post-Friedmannian formalism~\cite{HuSawicki2007a,Hu2008}, 
the Parametrized Post-Friedmannian (PPF) formalism~\cite{Skordis2008b,BakerEtAl2011,BakerFerreiraSkordis2013},
the effective fluid~\cite{BattyePearson2012,BattyePearson2013} and effective field theory formalisms~\cite{BloomfieldEtAl2012,GubitosiPiazzaVernizzi2012,PiazzaVernizzi2013}.~\footnote{Although \cite{HuSawicki2007a} and \cite{BakerFerreiraSkordis2013} use the same acronym, the two formalisms are widely different.}

Tests of gravity in the solar system have reached incredible precision.  
With the help of the PPN formalism, solar system tests such as lunar laser ranging and doppler tracking of the Cassini spacecraft
put bounds on the PPN parameters of around $10^{-3}-10^{-5}$ for curvature effects and $10^{-7} - 10^{-20}$ for
preferred frame effects~\cite{Will2014b}. However, the PPN formalism (as well as the PPE formalism) has as a basic tenet that the spacetime away from a source is 
asymptotically flat
%
%
and the gravitational theory remains perturbative down to the Schwarzschild radius of the source. But there are theories for which the latter assumption is manifestly untrue. Those theories have the property that the additional degrees of freedom to the metric become strongly coupled at some macroscopic scale, which in turn  makes it impossible to construct a perturbative expansion (as the PPN requires) that would be valid  from infinity all the way down to the Schwarzschild radius. But why should we bother with such theories?

The theories in question were constructed as a way to introduce departures from GR in the cosmological regime, i.e. in the limit of ultra-low curvatures and potentials. 
%
Whilst it is of academic interest to explore this possibility on purely theoretical grounds, most of these theories seek to address the cosmological  constant problem \cite{wein} and/or the origin of cosmic acceleration through a modification of gravity (for a review, see \cite{review}). Cosmic acceleration 
has by now been substantiated by a wide variety of observations.  Although hints for cosmic acceleration may be traced in studies of large scale structure in the 80's,
 the first real evidence came from measurements of the luminosity distance of type Ia supernovae  \cite{PerlmutterEtAl1998a,RiessEtAl1998}. 
Cosmic acceleration is thought to be caused by an effective fluid, called Dark Energy, whose nature is still a mystery.
The latest supernovae results from a combination of Sloan Digital Sky Survey (SDSS-II) and Supernovae Legacy Survey (SNLS) data constrain the equation of state of Dark Energy to
$w = P/\rho \simeq -1.018\pm0.057$ \cite{BetouleEtAl2014}, where $P$ is the dark energy pressure and $\rho$ its energy density.
 Other observational probes also indicate a Dark Energy component,
for instance, the latest measurements of the Cosmic Microwave Background (CMB)   anisotropies from the Planck Surveyor~\cite{AdeEtAl2013a} give $w = -1.13^{+0.13}_{-0.10}$
 while the cross-correlation between the Integrated Sachs-Wolfe (ISW) effect from the CMB and Large Scale Structure (LSS)
favour $w=-1.01^{+0.30}_{-0.40}$, at around $4\sigma$ \cite{GiannantonioEtAl2008}.

Attempts to explain cosmic acceleration by modifying the theory of gravity on cosmological scales ultimately require new degrees of freedom that make gravity behave rather differently from GR on ultra low curvatures. However, since the solar system data indicate that gravity is described by GR to a very good approximation 
in this regime, these degrees of freedom must somehow be hidden there. As solar system data (but also other data towards the strong gravitational field regime) are probing curvature scales much larger than cosmological curvatures this is entirely possible, at least observationally~\cite{BakerPsaltisSkordis2014}.
Indeed, this can happen if classical perturbation theory involving the new degrees of freedom breaks down at some large distance from a massive source, far beyond the Schwarzschild radius. In the case of the Sun, we typically require this large distance scale, known as the {\it Vainshtein radius}, to extend beyond the edge of the solar system, so that ``screening" occurs within the solar system itself  thanks to the nonlinear physics.
The mechanism by which this occurs 
is called the Vainshtein mechanism~\cite{Vainshtein1972,BabichevDeffayet2013,hirsute}, and is exploited by  so-called Galileon theories \cite{gal, bigal1,bigal2},  massive gravity \cite{mg1, mg2}, k-Mouflage gravity~\cite{BabichevDeffayetZiour2009},
 and the Fab Four \cite{f41,f42,f4nem}, 
to name just a few. An effective field theory for the Vainshtein mechanism has also been developed ~\cite{KoyamaNizTasinato2013} using the Horndeski action~\cite{Horndeski_orig,Generalized_Galileons}.
  However, the screening of these new degrees of freedom is not perfect, 
and some residuals can in principle be detected. This paper presents an extension of the PPN framework, the Parametrised Post-Newtonian Vainshteinian (PPNV) framework that 
is able to handle Vainshteinian corrections and thus
paving the way for determining the compatibility of such theories with solar system and other strong-field data. 
There has been some previous work in this direction, most notably \cite{clare,selfsc1,trod1,trod2,gustavo}.
%
%
Although it is easy to imagine how one can build a PPNV framework outside of the Vainshtein radius where classical perturbation theory remains valid, it is less obvious to see how to do this in the interior where the classical perturbative description has broken down.  Salvation lies in the so-called {\it classical dual} description \cite{GabadadzeHinterbichlerPirtskhalava2012,PadillaSaffin2012} , which ultimately corresponds to a Legendre transform of the original theory and admits a classical perturbative expansion {\it inside} the Vainshtein radiius (but not outside).  We will exploit this procedure in developing our framework. This initial work will focus on theories for which there is a unique macroscopic scale beyond the Schwarzschild radius at which point classical non-linearities start to kick in.

Whilst the need to develop a PPNV formalism is clearly important in testing these particular theories, we ought to issue a gentle word of warning about their validity once the non-lineariities begin to kick in.  The point  is that the break down of classical perturbation theory  is inherited from a breakdown in perturbative unitarity above some mass scale, $\Lambda$. From an effective field theory perspective, one would certainly expect there to be a tower of additional higher dimensional operators suppressed by the same scale $\Lambda$, included to preserve unitarity. As emphasised in \cite{unit}, such operators could, in principle, affect the classical solution out to macroscopic scales, possibly up to the Vainshtein radius itself. Without knowing the details of the (partial) UV completion beyond $\Lambda$, we do not know the impact of these operators but in principle they could affect the details of the PPNV expansions for a given theory both outside and inside the Vainshtein radius.  We emphasise that this paper seeks to build towards a PPNV framework for generic theories, which ultimately may be taken to include the aforementioned higher dimensional operators. However, for simplicity of presentation  we focus on the Cubic Galileon \cite{gal} as an illustrative  example, and do not worry about the precise details of which operators are present in that case, and that might have been omitted from a consistent UV completion.

The paper is organised as follows. We first describe the PPN framework, followed by a short description of the Vainshtein mechanism. We then outline how the PPN formalism has to be modified in order to include Vainshteinian corrections. As we will show further below, there are two separate perturbative regimes, one inside the Vainshtein radius and one outside, which must be handled separately. We then take the Cubic Galileon theory as an example and perform the calculation in the two regimes. We finish by considering spherical symmetry as a special case.

\section{Vainshteinian corrections to the PPN formalism: the PPNV expansion}

\subsection{The Standard Post-Newtonian Approximation}
The PPN formalism is a prescription for a perturbative expansion of the gravitational, matter and additional field equations of motion in successive orders of a small
parameter, the velocity of matter $v$ (in units of the speed of light). It was developed by Nordtvedt\cite{Nordtvedt1968} and 
later expanded refined by Will~\cite{Will1971} and Will and Nordtvedt~\cite{WillNordtvedt1972a}. Here we review the basic ingredients of PPN.

In PPN, one assumes that gravity is a geometric phenomenon and is described by a metric $g_{\mu\nu}$ and that matter only ``sees'' this metric $g_{\mu\nu}$ and follows its geodesics. 
Specifically this means that if $T_{\mu\nu}$ is the stress-energy tensor for matter  then
\begin{equation}
\nabla_\mu T^\mu_{\;\;\nu} = 0,
\label{T_cons}
\end{equation}
where $\nabla_\mu$ is the covariant derivative compatible with $g_{\mu\nu}$.
 However, the metric $g_{\mu\nu}$ may not be the only gravitational field. For instance,
there may be additional scalar fields to the metric that mediate the gravitational force (as is in our case), however, these fields do not enter the equations of motion for matter when those are formulated in the Jordan frame. Indeed, it is the so-called Jordan frame metric that should always be used in PPN.

To proceed, one first expands the metric $g_{\mu\nu}$ as a weak field perturbation $h_{\mu\nu}$  around Minkowski spacetime $\eta_{\mu\nu}$ as
\begin{equation}
\label{metricppnexpansion}
 g_{\mu\nu} = \eta_{\mu\nu}+h_{\mu\nu}.
\end{equation}
When doing explicit calculations, we shall adopt the convention that the metric has signature $(-, +, +, +)$.
The metric perturbation $h_{\mu\nu}$ is further expanded in successive orders dictated by a small parameter: the velocity of matter $v$ (we will assume units where the speed of light is unity).
Thus, the PPN is a small velocity expansion and is valid for gravitational fields generated by slowly moving matter.

The first correction to Minkowski spacetime is the Newtonian potential 
\begin{equation}
U(x)  \equiv \int\frac{\rho(\mathbf{x}',t)}{|\mathbf{x}-\mathbf{x}'|}d^3x'  \qquad \Leftrightarrow \qquad \grad^2 U = -4\pi \rho,
\label{defU}
\end{equation}
generated by the rest-mass density of matter $\rho$. This potential generates acceleration $ \mathbf{a} = G_N \grad U$ where $G_N$ is the measured Newton's constant by a lab experiment on the Earth.
Thus, virial relations dictate that $ v^2 \lesssim G_N U$, implying that the Newtonian potential has a PPN order $\O(2)$.
  In particular for spherical symmetry, we have that
$2 G_N U =  \frac{r_s}{r} \ll 1$ where $r_s = 2 G_N M$ is the Schwarzschild radius of a source with mass $M$. Moreover the acceleration equation gives the PPN order for derivatives, i.e.
spatial derivatives leave the PPN order unchanged while time derivatives have a PPN order $\frac{\partial}{\partial t} \sim \vec{v}\cdot \nabla \sim \O(1)$.
Of course, these considerations are not strict equalities and should be taken as a rule-of-thumb in order to assign PPN orders to the various quantities, such as $v$ or $U$ that may enter the equations of motion.
~\footnote{Spherically symmetric orbits around a point mass imply strict equality $v^2 = G_N U$ while a hypothetical body oscillating through the center of the Sun would have $v_{RMS}^2 \sim \frac{r_0^2}{R_{SUN}^2} G_N U_0$ where $r_0$ is the extend of its motion and $U_0$ is the gravitational potential at the centre. Hence once again $v_{RMS}^2 \lesssim G_N U$ as $r_0$ can be chosen at will. }

Other matter variables are also associated with a PPN order. The matter rest-mass density $\rho$ has a PPN order $\O(2)$ as is implied by (\ref{defU}). In addition to the density, matter sources may also have
specific energy density $\Pi$ 
(coming from other forms of energy, e.g. compressional, thermal, and corresponding to the ratio of the total of these energy densities  and the rest mass density of matter $\rho$) 
and pressure $P$, so that the matter stress-energy tensor takes the form 
\begin{equation}
 T_{\mu\nu}=(\rho+\rho\Pi+P)u_\mu u_\nu +  P g_{\mu\nu},
\label{defEMTensor}
\end{equation}
where the velocity four-vector is normalized as $g_{\mu\nu}u^\mu u^\nu=-1$. Since the pressure is generally smaller than the gravitational energy of matter ( $P \lesssim \rho U$ ) we associate it
with a PPN order $P \sim \O(4)$. Similar considerations apply to the specific energy density $\Pi \lesssim G_N U$ that is associated with a PPN order $\Pi \sim \O(2)$.
 The evolution of the matter fluid is determined by the continuity and Euler equations of motion that are obtained from (\ref{T_cons}) by expanding in PPN orders.

Clearly then, (\ref{defEMTensor}) implies that $\O(4)$ terms would appear in the (generalized) Einstein equations, thus we need to include these terms in the metric perturbation. The required order for the
metric perturbation components is found to be $h_{00} = h_{00}^{(2)} + h_{00}^{(4)}$  where the superscript number in the parenthesis denotes the PPN order of a particular term, 
$h_{0i} \sim \O(3)$ and $h_{ij} \sim \O(2)$~\cite{Will1981}. In the standard PPN formalism one has that $h_{00}^{(2)} = 2G_N U$, however, as we shall see further below, in the PPNV framework the Vainshteinian
corrections contribute to $h_{00}^{(2)}$ in the  form of new Vainshteinian potentials.  Furthermore, in the standard PPN gauge, one has $h_{ij} = 2 G_N \gamma U \delta_{ij}$ where $\gamma$ is one of the
PPN parameters. In the PPNV example we consider further below, one still has that $h_{ij} \propto \delta_{ij}$, however, $h_{ij}$ is no longer proportional to $h_{00}^{(2)}$ as in PPN due
to the appearance of new Vainshteinian potentials. We shall show that PPNV introduces new parameters that are not part of the standard PPN parameters.~\footnote{This is not a unique feature of the PPNV formalism. For instance in general bi-gravity theories a similar situation arises~\cite{CliftonBanadosSkordis2010}.}

In general, in PPN, one proceeds by determining the $R_{00}$ equation up-to $\O(4)$, the $R_{0i}$ equation  up-to $\O(3)$ and the $R_{ij}$ equation up-to $\O(2)$. Solving these equations in turn 
determines the metric to the required order and provides the definition of the PPN potentials. In addition to $U$, in the standard PPN gauge, there are a further $8$ PPN potentials
 that appear in orders higher than $2$.
Since in this article we are mostly concerned with the presentation of the PPNV extension compared to the PPN formalism, we shall not consider the $\O(3)$ and $\O(4)$ PPN orders, which are left to future work.
Therefore we will only need the $R_{00}$ and $R_{ij}$ equations  and the stress-energy tensor to $\O(2)$. These are
\begin{equation}
 R_{00}^{(2)} =    - \frac{1}{2} \grad^2  h_{00}^{(2)},    
\label{R_00_full}
\end{equation}
and
\begin{equation}
R_{ij} =  \frac{1}{2} \grad_k   \left[ \grad_i h^k_{\;\;j} + \grad_j h^k_{\;\;i}\right]
 - \frac{1}{2}  \grad^2 h_{ij}
 + \frac{1}{2} \grad_i \grad_j (  h_{00}^{(2)} - h),
\label{R_ij_full}
\end{equation}
where $h = h^i_{\;\;i}$.  We will also need $T_{00} = \rho$ and the trace $T = T^\mu_{\;\;\mu} = -\rho$.
Let us now discuss the Vainshtein mechanism and the modifications it introduces, leading to the PPNV expansion.

\subsection{Vainshtein Mechanism for a Spherically Symmetric Source}
\label{sec_spherical}
 As we already discussed in the previous subsection,  the metric theories treated with the standard PPN formalism (e.g. GR or Brans-Dicke~\cite{Will1981}) 
 have a single expansion parameter given by $v^2 \sim \frac{r_s}{r}$. In contrast, the non-linearities of the scalar field 
 in theories with a Vainshtein mechanism (such as the Galileon theories) introduce new regimes that must be handled along with PPN. 
These Vainshteinian thresholds are associated with an extra expansion parameter $\frac{r}{r_V}$ that accounts 
for new physical effects associated with the new scale introduced by these theories, the Vainshtein scale $r_V$.

  A simple way to think about their different expansions is the following. The ratio $\frac{r_s}{r}\ll 1$ accounts for the non-relativistic nature of the matter
 sources considered in the PPN approximation while the ratio $\frac{r}{r_V}$ determines whether the observer is placed either inside or outside the Vainshtein
 radius $r_V$ and how far from the source (or close to $r_V$) and consequently how important the Vainshteinian corrections are. Clearly, we must then consider two regimes: inside the Vainshtein
radius so that $\frac{r}{r_V} \ll 1$ and outside the Vainshtein radius where $\frac{r}{r_V} \gg 1$. These two regimes must be treated separately.

As an example consider static spherically symmetric solutions in the Cubic Galileon theory in the Einstein frame (see Eq. \ref{galact}). 
The metric $g_{\mu\nu}$ that enters (\ref{T_cons}) is a combination of the Einstein metric $\tilde{g}_{\mu\nu}$ and the scalar field $\chi$ (assumed dimensionless)
as $g_{\mu\nu} = e^{2\chi} \tilde{g}_{\mu\nu}$. Thus, considering weak fields on Minkowski spacetime the effective gravitational potential would be
$h_{00} = \tilde{h}_{00} - 2 \chi = \frac{2G M}{r} - 2 \chi$.
The scalar equation of motion for a spherically symmetric source of mass $M$ centered at $r=0$ is 
\begin{equation}
\frac{2\omega+3}{r^2} \frac{d}{dr}\left[r^2 \chi'\right] + \frac{\alpha}{r^2}\frac{d}{dr}\left[ r (\chi')^2\right] =  G M  \frac{\delta(r)}{r^2},
\label{phispherical_2nd}
\end{equation}
where 
\begin{equation}
\alpha = \frac{M_p}{\Lambda^3}.
\label{def_alpha}
\end{equation}
Now (\ref{phispherical_2nd}) can be integrated once to give
\begin{equation}
 (2\omega+3) r^2 \chi'+ \alpha  r(\chi')^2   =  G M.
\end{equation}
Thus, solving for $\chi'$ we find
\begin{equation}
\chi' =  \frac{2\omega+3}{2\alpha}  r \left[ - 1 + \sqrt{ 1  + \frac{4 G M \alpha }{  (2\omega+3)^2 r^3  }   } \right],
\label{chi_p_spher_sym}
\end{equation}
where the plus sign in front of the square root is chosen so that no divergence appears when $\alpha \rightarrow 0$.
 There are two limiting cases. As $\alpha \rightarrow 0$,
or in other words, when $ \frac{4  G M \alpha }{  (2\omega+3)^2 r^3  }   \ll 1$ then 
\begin{equation}
\chi =  
- \frac{ G M  }{ (2\omega+3) r  } 
  + \frac{  (G M)^2 }{ 4 (2\omega+3)^3   } \frac{\alpha}{r^4}
 + \ldots
\end{equation}
In the opposite limit as $\alpha \rightarrow \infty$, or in other words, when $ \frac{4 G M \alpha }{  (2\omega+3)^2 r^3  }   \gg 1$ then
\begin{equation}
\chi =  2\sqrt{\frac{ G M }{\alpha}}    r^{1/2},
\end{equation}
where the integration constant is ignored as it can be rescaled away by a coordinate transformation.

The turning point occurs when $ \frac{4 G  M  \alpha}{ (3+2\omega)^2 r^3  } = 1$, i.e. at a radius $\frac{1}{(2\pi)^{1/3}(3+2\omega)^{2/3}}  r_V$ 
where $r_V$ is the scale  given by
\begin{equation}
 r_V = \frac{1}{\Lambda} \left[ \frac{M}{M_p} \right]^{1/3}.
\label{rv_gal}
\end{equation}
This $r_V$ is called the Vainshtein scale and will play a fundamental role in what follows. 

The Vainshtein mechanism is now clear. For small distances away from the source, $r\ll r_V$, the effective gravitational potential 
is 
\begin{equation}
h_{00} = \frac{r_s}{r} \left[1    -  4 \sqrt{ 2 \pi}   \left(\frac{r}{r_V}\right)^{3/2}  + \ldots \right],
\end{equation}
 while for large distances away from the source, $r\gg r_V$, it is 
\begin{equation}
h_{00} = \frac{\tilde{r}_s}{r} \left[ 1
  - \frac{1}{ 64\pi (2\omega+3)^2 (2 + \omega)  }  \left(\frac{r_V}{r}\right)^3
 + \ldots
\right]
\end{equation}
 where $\tilde{r}_s= \frac{4+2\omega}{3+2\omega} r_s$ is a rescaled Schwarschild radius.
Note that with $r_V$ now defined, the $\alpha$ parameter can be written as $\alpha  \sim \frac{ r_V^3 }{r_s} $

We saw by example what we  should expect in a theory that leads to a Vainshtein mechanism. In the next subsection we formalise the treatment above
and consider general theories with a Vainshtein mechanism. In particular we shall see how the Vainshtein scale and the expansion parameter 
$\alpha$ emerge, and what impact this can have on the PPNV expansion.
 
\subsection{The PPNV formalism} \label{sec:PPNV}
To develop a Parametrised Post-Newtonian Vainshteinian (PPNV) formalism, we must first understand some 
of the generic features of modified gravity theories with Vainshtein screening. The modification of gravity is normally associated with a 
light scalar degree of freedom that strongly affects the dynamics on cosmological scales when linearised perturbation theory is valid.  
However, in the presence of a heavy source, linearised perturbation theory  breaks down at some macroscopic scale owing to derivative interactions involving the new field $\chi$. Let us illustrate this with a generic example, schematically described by an action in the Einstein frame\footnote{If gravity is modified  by a long range scalar, you would expect the leading order canonical kinetic structure presented here in any sort of sensible  Lorentz invariant set up that is weakly coupled at long distances.}: 
\begin{equation} 
S \sim \int  d^4x \left\{M_p^2\sqrt{-g} R- (\del \chi)^2+O(\chi) + M_p h_{\mu\nu}\bar O^{\mu\nu}(\chi)+h_{\mu\nu}T^{\mu\nu}+\frac{\chi}{M_p} T
\right\},
\end{equation}
where $M_p^2 = \frac{1}{8\pi G}$ and where in general $G/G_N \sim \O(1)$ (usually the measured Newtonian strength $G_N$ is not equal to the bare gravitational
strength $G$ in the action but is typically related to it by an $\O(1)$ quantity).
This schematic form encompasses a number of modified gravity scenarios in the so-called decoupling limit, including single field Galileons \cite{gal} and dRGT massive gravity\footnote{In massive gravity, the scalar $\chi$ is identified with the Stuckelberg scalar.The vector modes are subleading in the decoupling limit, so our method would have to be adapted to include them. } \cite{mg1,mg2}.  The canonical scalar field is coupled directly to matter with gravitational strength, but the Vainshtein mechanism is made possible by the derivative interactions between the scalar and the graviton and/or the self interactions. Assuming for the moment a unique strong coupling scale, $\Lambda$, we have that these operators contain terms like
\begin{equation} 
 \label{int}
  O(\chi) \supset \frac{\del^{2m} \chi^n}{\Lambda^{2m+n-4}}, \qquad \bar O^{\mu\nu}(\chi) \supset \frac{\del^{2\bar m} \chi^{\bar n-1}}{\Lambda^{2\bar m+\bar n-4}},
\end{equation} 
where the integers $m, \bar m \geq 1, ~n, \bar n \geq 3$. The strong coupling scale marks the breakdown of perturbative unitarity, and in principle one may have to include a whole tower of higher dimensional operators to preserve unitarity at higher energies.  

For a static and spherically symmetric profile, the classical potentials  at large distances from a  heavy  source of mass $M$ are simply Newtonian, $M_p h^c_{\mu\nu} \sim \chi_{lin}^c \sim \frac{M}{M_p}\frac{1}{r}$. Assuming this Newtonian behaviour, the interactions in (\ref{int}) would become comparable with the leading order canonical terms when 
$$
\frac{1}{\Lambda^{2m+n-4}} \frac{1}{r^{2m}}\left(\frac{M}{M_p}\frac{1}{r} \right)^n \sim \frac{1}{r^2} \left(\frac{M}{M_p}\frac{1}{r}\right)^2 \implies r \sim \frac{1}{\Lambda}\left(\frac{M}{M_p}\right)^{\frac{n-2}{2m+n-4}},
$$
with similar expressions for the barred integers.  We see that in general different interactions yield different macroscopic scales. As a first pass in developing the post-Vainshteinian formalism, let us only consider scenarios in which there is a unique macroscopic scale beyond the Schwarzschild radius, so that all nonlinear interactions become important at the same scale. In other words, we only consider those theories for which there is a unique strong coupling scale, $\Lambda$, and for which all interactions of the form (\ref{int}) yield the same value of $\frac{n-2}{2m+n-4}=s, \frac{\bar n-2}{2\bar m+\bar n-4}=s$. An example of this would be a  combination of Galileon interactions suppressed by the same strong coupling scale, each giving $s=\frac13$. Therefore, we shall begin to develop our formalism only for those theories for which there is a unique macroscopic scale of interest, which we then identify with a unique Vainshtein radius, 
\begin{equation}
r_V \sim \frac{1}{\Lambda} \left(\frac{M}{M_p}\right)^s,  \qquad s=\frac{n-2}{2m+n-4}= \frac{\bar n-2}{2\bar m+\bar n-4}; \forall n,m \in O(\chi), \forall \bar n, \bar m \in \bar O^{\mu\nu}(\chi),
\label{rv}
\end{equation}
 signaling the breakdown of classical perturbation theory. This is how the strong coupling manifests itself classically. Generalisations that take care of multiple macroscopic scales ( i.e. interactions with different strong coupling scales, and different values of $s$) and multiple expansion parameters will be left for future study.

%
%

Let us recall that in the PPN formalism the leading order contribution is the Newtonian potential, satisfying $\nabla^2 U=- 4\pi G_N \rho$ and is defined to be  $\O(2)$ in PPN.  The orders of velocities and time derivatives are then inferred using the virial  ($v^2 \sim U$) and Euler ($\del_t \sim v \cdot \nabla$) relations.  In the spherically symmetric scenario described above, we note that $U \sim \frac{r_s}{r}$, where $r_s\sim M/M_p^2$ is the Schwarzschild radius of the source, so there is a sense in which the PPN expansion is an expansion in $\sqrt{ \frac{r_s}{r}}$.  Similarly, we shall think of PPV as an expansion in $\left(\frac{r}{r_V}\right)^\frac32$. Although our analysis here is generic, the exponent $\frac32$ is ultimately motivated by the Cubic Galileon scenario. This is a sensible starting point since the Cubic Galileon  is the lowest dimensional operator corresponding to a purely derivative self-interaction of a single scalar in a Lorentz invariant theory.


To gain further insight, consider what happens deep inside the Vainshtein radius. If the scalar dynamics is dominated by a pure interaction term, we would have
$$
 \frac{\del^{2m}( \chi_{nonlin}^c)^{n-1}}{\Lambda^{2m+n-4}} \sim \frac{T}{M_p},
$$
for some $m, n$, and so for a static spherically symmetric configuration,  
$$
\frac{\chi_{nonlin}^c}{M_p} \sim \left(\frac{r}{r_V}\right)^\frac{2m+n-4}{n-1} \left(\frac{r_s}{r}\right).
$$
This suggests that the scalar part of the physical metric that couples to matter is $\O(2)$ in PPN and $\O\left(\frac23\left[\frac{2m+n-4}{n-1}\right]\right)$in PPV. However, if the deep Vainshteinian behaviour is dominated by a mixed interaction then we would schematically have
$$
M_p h_{\mu\nu} \frac{\del^{2\bar m}( \chi_{nonlin}^c)^{\bar n-2}}{\Lambda^{2\bar m+\bar n-4}} \sim \frac{T}{M_p},
$$
for some $\bar m, \bar n$. For  a static spherically symmetric configuration in which screening is effective, we expect $M_p \del^2 h_{\mu\nu} \sim \frac{T}{M_p}$, and so
$$
\frac{\chi_{nonlin}^c}{M_p} \sim \left(\frac{r}{r_V}\right)^\frac{2\bar m+\bar n-4}{\bar n-2} \left(\frac{r_s}{r}\right).$$ This now suggests that the scalar part of the physical metric that couples to matter is $\O(2)$ in PPN and $\O\left(\frac23\left[\frac{2\bar m+\bar n-4}{\bar n-2}\right]\right)$ in PPV. 

The previous paragraph indicates two important things: (i) that even in the deep Vainshteinian region the leading order behaviour of the physical metric is Newtonian, as it should be; and (ii) that what we really have here is a double expansion, owing to the hierarchical difference between the the Vainshtein radius and the Schwarzschild radius. To this end we assign a combined order $\O(N,V)$ to a quantity that is order $N$ in PPN, and order $V$ in PPV.  In terms of Schwarzschild and Vainshtein radii, we may think of $$\O(N,V) \sim  \left(\frac{r_s}{r}\right)^\frac{N}{2} \left(\frac{r}{r_V}\right)^\frac{3V}{2}.$$
Thus, $U$ and $\chi^c_{lin}$  are $\O(2, 0)$, whilst $\chi^c_{nonlin}$ is $\O\left(2, \frac23\left[\frac{2m+n-4}{n-1}\right[\right)$ or $\O\left(2, \frac23\left[\frac{2\bar m+\bar n-4}{\bar n-2}\right]\right)$, depending on whether it is a pure scalar, or a mixed interaction that dominates the deep Vainshteinian region.

PPN can be thought of as an expansion in the Schwarzschild radius, and is therefore equivalent to an expansion in the source. Similarly  PPNV can be thought of as an expansion in Schwarzschild and Vainshtein radii, and for practical purposes this is best realised in terms of an expansion in the source, and an operator, $\alpha$,  that is independent of the source, but which carries PPV order\footnote{This perspective is a bit sloppy in both PPN and PPNV: the source, $\rho$, is not dimensionless and therefore one might consider it a poor expansion parameter. In PPNV, the same may be said of $\alpha$ given by (\ref{alpha}). However, we remind the reader that what is really going on is an expansion in $r_s/r$ and $r/r_V$, with $\rho$ and $\alpha$ simply being a more convenient way of keeping track of that expansion. }.  Given the Vainshtein radius (\ref{rv}), there is only one candidate for $\alpha$
 (or powers thereof), namely:
\begin{equation} \label{alpha}
 \alpha \sim M_p/\Lambda^{1/s}.
\end{equation}
  It is easy to see  that $\alpha $ is $\O\left(-2,-\frac{2}{3s}\right)$.  Operators containing powers of $\Lambda$ should now be rewritten in terms of $\alpha$. We will also assume that velocities are always determined by the virial relation with the Newtonian potential, since the scalar potential never dominates over the (Einstein frame) graviton if  screening is active.  The rules of the game are therefore as follows:  the orders for velocities ($v \sim \O(1,0)$), time derivatives ($\del_t \sim \O(1,0)$), space derivatives $\del_i \sim \O(0,0)$, total energy density $\rho \sim  \O(2,0)$, specific energy density ($\Pi \sim \O(2,0)$), and pressure $P\sim \O(4, 0)$ are inherited from PPN \cite{Will1981}, and do not care about the PPV expansion. The only operator to carry PPV order now is $\alpha \sim \O\left(-2,-\frac{2}{3s}\right)$. 

For a given theory, we write the action explicitly in terms of $\alpha$, and study the resulting field equations order by order in the double expansion. Schematically, for our generic example, we have
\begin{equation}
 \label{actschem}
S \sim \int d^4x \left\{ M_p^2\left[\sqrt{-g} R- (\del \chi)^2+P(\chi)+h_{\mu\nu}\bar P^{\mu\nu}(\chi)\right]+h_{\mu\nu}T^{\mu\nu}+\chi T
\right\},
\end{equation}
where we have rescaled $\chi\to M_p \chi$, and
\begin{equation} 
\label{P}
P(\chi)=\frac{O(M_p \chi)}{M_p^2} \supset \alpha^{n-2}  \del^{2m} \chi^n, \qquad \bar P^{\mu\nu}(\chi)=\frac{ \bar O^{\mu\nu}(M_p \chi)}{M_p} \supset  \alpha^{\bar n-2}  \del^{2\bar m} \chi^{\bar n-1}.
\end{equation}
Recall that we have assumed each interaction has the unique value of  $s=\frac{n-2}{2m+n-4},   \frac{\bar n-2}{2\bar m+\bar n-4}$ where $n,m \in O(\chi), \bar n, \bar m \in \bar O^{\mu\nu}(\chi)$. The PPNV analysis is straightforward beyond the Vainshtein radius where the leading order contributions are purely Newtonian i.e. zeroth order in PPV, and corrections carry {\it negative} PPV order as higher order terms in $ \alpha$ begin to kick in. These corrections generate new classes of PPV potentials, and a complete formalism should ultimately include all possibilities, or at the very least, all possibilities that are relevant to models that appear in the literature. 

In contrast, deep inside the Vainshtein radius, the leading behaviour should again be Newtonian but with corrections carrying {\it positive} PPV order, requiring  terms proportional to inverse powers of $\alpha$. However, the action expressed in (\ref{actschem}) is not well suited to an expansion in terms of these inverse powers. 
This is where the so-called {\it classical dual} \cite{GabadadzeHinterbichlerPirtskhalava2012,PadillaSaffin2012} comes in. 
For our schematic action (\ref{actschem}), there exists a dual action describing the same classical physics \cite{PadillaSaffin2012}
\begin{multline} \label{dualschem}
S \sim \int  M_p^2\left\{\sqrt{-g} R- (\del \chi)^2+\sum_{r =0}^{r_\text{max}}\left[ \frac{\del F }{\del A_{\mu_1 \ldots \nu_r}}+ h_{\alpha\beta} \frac{\del \bar F^{\alpha\beta} }{\del A_{\mu_1 \ldots \nu_r}}\right] \nabla_{i\mu_1} \ldots \nabla_{\mu_r} \chi \right. \\ \left.+\alpha^{-s t} \left[ F- \sum_{r}^{r_\text{max}} \frac{\del F }{\del A_{\mu_1 \ldots \mu_r}}A_{\mu_1 \ldots \mu_r} \right] +\alpha^{-s t}  h_{\alpha\beta} \left[ \bar F^{\alpha\beta} - \sum_{r}^{r_\text{max}}\frac{\del \bar F^{\alpha\beta}   }{\del A_{\mu_1 \ldots \mu_r}}A_{\mu_1 \ldots \mu_r} \right]\right\}+h_{\mu\nu}T^{\mu\nu}+\chi T,
\end{multline}
where $r_{\text{max}}$ is the largest value of $2m$, and $t={\text{max}\left\{ \frac{2m+n-4}{n-1},   \frac{2\bar m+\bar n-4}{\bar n-2}; n,m \in O(\chi), \bar n, \bar m \in \bar O^{\mu\nu}(\chi) \right\}}$.  We also have combinations of polynomials in the auxiliary fields $A_{\mu_1 \ldots \mu_r}$, i.e. 
\begin{equation}
F \supset F_n(A, A_\mu, \ldots), \qquad \bar F^{\alpha\beta} \supset \bar F^{\alpha\beta}_{\bar n-1}(A, A_\mu, \ldots),
\end{equation}
where $F_n$ is order  $n$ in the auxiliary fields, while $\bar F^{\alpha\beta}_{\bar n-1}$ is order $(\bar n-1)$.  These polynomials stem from the interactions given in  (\ref{P}). Introducing 
\begin{eqnarray}
\delta(m,n) &=&   \frac{n-1}{2m+n-4}\left(t-\frac{2m+n-4}{n-1}\right)\geq 0  \qquad  \forall n,m \in O(\chi), \\
\bar \delta (\bar m, \bar n) &=& \frac{\bar n-2} {2\bar m+\bar n-4}\left(t-\frac{2\bar m+\bar n-4}{\bar n-2}\right)\geq 0 \qquad \forall \bar n, \bar m \in \bar O^{\mu\nu}(\chi) ,
\end{eqnarray}
it can be shown that $F_n$ scales as $\alpha^{-\delta \left(n-2 \right)}$, while $\bar F^{\alpha\beta}_{\bar n-1}$ scales as $\alpha^{-\bar \delta (\bar n-2) \left(\bar \epsilon+1\right)}$. We now have an action made up of non-negative powers of $1/\alpha$, and as such it is well suited to an expansion that increases both the PPV order (as desired) and the PPN order. The PPV expansion deep inside the Vainshtein radius generates yet another class of PPV potentials, and due to the non-linear nature of the problem to leading order, we do not expect to be able to write these in closed form. 


 Let us now summarize the steps that need to be followed in order to apply the
PPNV expansion for the class of theories under consideration: i.e. those for which all interactions kick in at the same macroscopic scale. 
\begin{itemize}
   \item Determine the Vainshtein radius $r_V$ for the theory. This is usually determined from the Einstein frame action using (\ref{rv}) and the discussion that precedes it.
   \item Form the expansion parameter $\alpha$ and  determine its PPNV order.
   \item For the outside region, reduce the field equations of the theory {\it in the Jordan frame}, 
     as is done in the case of PPN. Note that $\alpha$ usually carries a PPN order, so care must be taken to 
     take this into consideration when performing the expansion. 
    \item For each PPN order, solve the field equations to determine which potentials arise. Furthermore, for each PPN order, solve the scalar equation in PV orders to determine the metric
potentials to the required Vainshtein order.
   \item Dualize the theory in the Jordan frame and determine the field equations. 
   \item For the inside region, reduce the field equations of the theory {\it in the Jordan frame}, 
     as is done in the case of PPN. 
   \item For each PPN order, solve the field equations to determine which potentials arise. Furthermore, for each PPN order, solve the scalar equation in PV orders to determine the metric
potentials to the required Vainshtein order.
\end{itemize}
We shall now develop the PPNV scheme for Cubic Galileon theory in order to illustrate the idea.

\section{Case study: Cubic Galileon theory}
\subsection{A short introduction to Cubic Galileon theory}
In order to demostrate our formalism, we shall apply it to a particular case of a screening theory: Cubic Galileons.
We begin by presenting the action and field equations in the standard form (we shall give an alternative, dual formulation further below). The action in the Einstein frame is
\begin{equation} \label{galact}
S[\tilde{g},\chi] = \frac{1}{16\pi G} \int d^4x \sqrt{-\tilde{g}}  \tilde{R} + \int d^4x \sqrt{-\tilde{g}}  \left(c_0 X 
 +  \frac{1}{\Lambda^3} X \tilde{\square} \chi\right) + S_M[g],
\end{equation}
where $G = \frac{1}{8\pi M_p^2}$ is the bare gravitational strength, $c_0>0$ is a constant, $X  = - \frac{1}{2} \tilde{g}^{\mu\nu} \nabla_\mu \chi \nabla_\nu \chi$
and $\Lambda$ is the strong coupling scale when $c_0 \sim \O(1)$. $G$ and $M_p$ are not the measured value of Newton's constant and Planck mass but are rather parameters in the action; we shall return to this point further below.

Following the discussion above (\ref{rv}), we determine the Vainshtein scale. We have that for the Cubic Galileon the $\bar{O}^{\mu\nu}$ term is absent while the 
$O(\chi)$ term is $  \frac{1}{\Lambda^3} X \tilde{\square} \chi$. Hence, $m = 2$ and $n= 3$ which gives the Vainshtein scale as $r_V = \frac{1}{\Lambda} \left(\frac{M}{M_p}\right)^{1/3}$
as in (\ref{rv_gal}). Having determined the Vainshtein scale, we also form the expansion parameter $\alpha = \frac{M_p}{\Lambda^3}$. The expansion orders for $\alpha$ are then found to be
$\alpha \sim \O(-2,-2)$, in other words, $\alpha$ lowers both the PPN and the Vainshtein order of any terms multiplying it, by $2$.

The Jordan frame is determined via a conformal transformation to a new metric $g_{\mu\nu} =  e^{2\chi/M_p} \tilde{g}_{\mu\nu}$ that
is minimally coupled to matter. Defining $\phi =  e^{-2\chi/M_p}$ the action in the Jordan frame takes the form
\begin{equation}\label{actionSCG}
S = \frac{1}{16\pi G}\left\{\int d^4x \sqrt{-g}\left[ \phi R +\frac{2\omega}{\phi}Y \right]  -\frac{\alpha}{4}  \int d^4x \sqrt{-g}  \frac{Y}{\phi^3}  \square\phi 
\right\} +S_m[g,\psi^A],
\end{equation}
where $\omega = \frac{c_0-6}{4}$  and  $ Y \equiv - \frac{1}{2} g^{\mu\nu} \nabla_\mu \phi \nabla_\nu \phi$. 
The matter action $S_m$ depends, in addition to a generic set of matter fields $\psi^A$, only on the metric $g_{\mu\nu}$ but not on the Galileon field $\phi$.

Varying the action with respect to $g_{\mu\nu}$ and $\phi$ and after some algebraic manipulations gives the generalized Einstein equations as
\begin{eqnarray}
 \phi  R_{\mu\nu}  &=& 8 \pi G \left[ T_{\mu\nu} - \frac{1}{2} T g_{\mu\nu} \right]
+     \frac{\omega}{\phi}  \nabla_\mu \phi  \nabla_\nu \phi   
 + \frac{1}{2} \square \phi g_{\mu\nu}  +    \nabla_\mu \nabla_\nu \phi
\nonumber 
\\
&&
+ \frac{\alpha}{8\phi^3}  \bigg\{
- Y     \square\phi  g_{\mu\nu}
+  \left[ 6 \frac{Y}{\phi}     -  \square\phi \right] \nabla_\mu \phi  \nabla_\nu \phi
   - \nabla_\mu Y \nabla_\nu \phi - \nabla_\mu \phi \nabla_\nu Y  
\bigg\},
 \label{EinsEq1} 
\end{eqnarray}
and scalar equation as
\begin{equation}
 (3 + 2 \omega )    \square \phi 
+ \frac{   \alpha }{4\phi^2} \bigg[
 \frac{5}{\phi}   \nabla_\mu \phi  \nabla^\mu Y 
+ \frac{18Y^2}{\phi^2} 
-     (\square\phi)^2
-   \nabla^\mu \phi  \nabla_\mu   \square\phi 
-      \square Y 
 -  Y   \frac{1}{\phi}  \square\phi 
\bigg]
= 
 8 \pi G T.
 \label{ScalarEq} 
\end{equation}

We have chosen this form of the (generalized) Einstein equations as it is more convenient when applying the PPN (and by extension the PPNV) formalism.

\subsection{The PPNV expansion outside the Vainshtein radius}
Outside the Vainshtein radius the self-interactions of the Galileon are subdominant and the matter terms lead the sources of the field equations. 
It is straightforward to see
that, at leading order the field equation (\ref{ScalarEq}) basically reduces to that of a Brans-Dicke field. In contrast to what happens inside the Vainshtein region, 
for sufficiently large scales the Galileon shows up as a linear field with small corrections due to the scalar self-interactions that may be responsible for modifications 
to the cosmological dynamics. 

Because the source is assumed to be non-relativistic it is reasonable to split the field in PPN orders

\begin{equation}
 \phi= \phi_0^{(out)}(1 + \varphi^{(2)} + \varphi^{(4)}+...),
\label{phippnexp}
\end{equation}
 where the superscript denotes the PN order of each term and $\phi_0^{(out)}$ is the constant background (cosmological) value of the field outside the Vainshtein radius.
We now proceed to the generalized Einstein equations (\ref{EinsEq1}) order-by-order.

\subsubsection{$h_{00}$ to $\O(2)$}

Using (\ref{R_00_full}) the (generalized) $00$ Einstein equations outside the Vainshtein radius give at order $\O(2)$ 
\begin{eqnarray}
   - \frac{1}{2} \grad^2  h_{00}^{(2)}    &=& 4 \pi \tilde{G}  \rho- \frac{1}{2}   \grad^2 \varphi^{(2)},
\label{eq_R_00_out}
\end{eqnarray}
where we have defined
\begin{equation}
  \tilde{G} = \frac{G}{ \phi_0^{(out)} }.
\end{equation}
 Notice that the above equation   justifies $\varphi \sim \O(2) + \ldots$.
The solution to (\ref{eq_R_00_out}) is 
\begin{equation}
 h_{00}^{(2)} = 2 \tilde{G}  U + \varphi^{(2)},
\label{eq_h_00_2_standard}
\end{equation}
where
$U$ is the standard Newtonian potential (\ref{defU}).

At this point, the solution (\ref{eq_h_00_2_standard}) is formally identical to the solution for $h_{00}$ in Brans-Dicke theory. However,
 we do not yet know the solution for the new potential $\varphi^{(2)}$, which will come out to be different than the Brans-Dicke case. 

\subsubsection{$h_{ij}$ to $\O(2)$}
Now consider the $R_{ij}$ equation to $\O(2)$. 
Using  (\ref{R_ij_full}) in (\ref{EinsEq1}) we find
\begin{equation}
 \frac{1}{2} \grad_k   \left[ \grad_i h^k_{\;\;j} + \grad_j h^k_{\;\;i}\right]
 - \frac{1}{2}  \grad^2 h_{ij}
 + \frac{1}{2} \grad_i \grad_j (  h_{00}^{(2)} - h)
  = 4 \pi \tilde{G}\rho \gamma_{ij} + \frac{1}{2} \grad^2 \varphi^{(2)}  \gamma_{ij} +   \grad_i \grad_j \varphi^{(2)}.
\label{eq_R_ij_standard_1}
\end{equation}
We impose the gauge-fixing condition
\begin{equation}
 \grad_k h^k_{\;\;i} =  \grad_i \left( \frac{1}{2} h -  \frac{1}{2} h_{00}^{(2)}  + \varphi^{(2)} \right),
\label{eq_GF_standard_1}
\end{equation}
which brings (\ref{eq_R_ij_standard_1}) into the form
\begin{equation}
 - \frac{1}{2}  \grad^2 h_{ij} = 4 \pi \tilde{G}\rho \gamma_{ij} + \frac{1}{2} \grad^2 \varphi^{(2)}  \gamma_{ij}.
\label{eq_R_ij_standard_2}
\end{equation}
The solution to the above equation is
\begin{equation}
h_{ij} = \left[ 2 \tilde{G} U  - \varphi^{(2)} \right] \gamma_{ij}.
\label{eq_h_ij_standard}
\end{equation}

With (\ref{eq_h_00_2_standard}) and (\ref{eq_h_ij_standard}) we have now determined the metric to $\O(2)$ in PPN in terms of the Newtonian potential $U$ and the still unknown field
$\varphi^{(2)}$. To proceed further we must now consider the scalar equation to $\O(2)$.

\subsubsection{Scalar equation to $\O(2)$}
The scalar equation to  $\O(2)$ gives
\begin{equation}
\grad^2 \varphi^{(2)} - \frac{\alpha }{4(3+ 2\omega)\phi_0^{(out)}} \bigg\{ 
 (\grad^2 \varphi^{(2)} )^2
+   \grad \varphi^{(2)} \cdot  \grad  \grad^2 \varphi^{(2)}
-\frac{1}{2}  \grad^2  |\grad \varphi^{(2)}|^2
\bigg\}
= 
 -\frac{8 \pi \tilde{G}}{3+2\omega}  \rho.
\label{eq_phi_O2}
\end{equation}
Taking $\alpha\rightarrow 0$ recovers the Brans-Dicke equation for the scalar that can be easily solved.
In the general $\alpha\ne 0$ case, however, it is impossible to solve the above equation analytically, except in idealized situations, for instance, spherical symmetry. In order to make progress we appeal to the PPNV expansion: we can find the solution by expanding  in Vainshtein orders as
\begin{equation}
 \varphi^{(2)} =  \varphi^{(2,0)} +  \varphi^{(2,-2)}+  \varphi^{(2,-4)} + \ldots =  \sum_{n=0}^{\infty}\varphi^{(2,-2n)}.
\end{equation}

\subsubsection{Metric solution to  $\O(2,0)$}
Consider first the leading Vainshtein order, i.e. $\O(2,0)$. In this case we recover the  Brans-Dicke equation that is solved as
\begin{equation}
 \varphi^{(2,0)} = \frac{2\tilde{G}}{3+2\omega}U.
\label{phi20}
\end{equation}
Therefore to leading order we find using (\ref{eq_h_00_2_standard}) and (\ref{eq_h_ij_standard})
\begin{eqnarray}
  h_{00}^{(2,0)} &=& 2 G_C U,
\\
  h_{ij}^{(2,0)} &=&  2 G_C \gamma \; U  \; \gamma_{ij},
\end{eqnarray}
where we have defined the ''cosmological'' gravitational strength $G_C$ as
\begin{equation}
  G_C =  \frac{4+2\omega}{3+2\omega} \tilde{G},
\end{equation}
and recovered the PPN parameter $\gamma$ for Brans-Dicke theory:
\begin{equation}
\gamma = \frac{1+\omega}{2+\omega}.
\end{equation}
Thus to leading order, the Cubic Galileon theory reduces to Brans-Dicke theory outside the Vainshtein radius. In fact, it can be shown that this feature is valid in the PPN expanstion to $\O(4)$, 
however, due to the more complicated equations in that case we leave it for a separate publication.

\subsubsection{Metric solution to  $\O(2,-2)$}
We now solve the scalar equation to $\O(2,-2)$. Keeping on the the $\O(2,-2)$ terms, the scalar equation (\ref{eq_phi_O2}) gives
\begin{eqnarray}
&&
\grad^2 \varphi^{(2,-2)} 
= 
 \frac{   \alpha }{8(3+ 2\omega)\phi_0^{(out)}} \bigg\{ 
2\grad \left[\grad \varphi^{(2,0)} \grad^2 \varphi^{(2,0)} \right]
-  \grad^2  |\grad \varphi^{(2,0)}|^2
\bigg\}.
\end{eqnarray}
Noting the Laplacian acting on the 2nd term we define
\begin{equation}
 \varphi^{(mix)} = \varphi^{(2,-2)} + \frac{   \alpha }{8(3+ 2\omega)\phi_0^{(out)}}   |\grad \varphi^{(2,0)}|^2,
\end{equation}
so that our equation after using (\ref{phi20}) becomes
\begin{eqnarray}
&&
\grad^2 \varphi^{(mix)} 
= 
- \frac{ \alpha }{4(3+ 2\omega)\phi_0^{(out)}}  \frac{8 \pi \tilde{G}}{3+2\omega}   \grad \cdot \left[\rho \grad \varphi^{(2,0)}  \right].
\end{eqnarray}
Then we can solve the above equation to get
\begin{equation}
 \varphi^{(mix)}(t,\vec{x})
= 
 \frac{ \alpha }{4(3+ 2\omega)\phi_0^{(out)}} \frac{2 \tilde{G}}{3+2\omega}   \int d^3x' \frac{1}{|\vec{x}-\vec{x}'|} 
\grad_{x'} \cdot \left[\rho(t,\vec{x}') \grad_{x'} \varphi^{(2,0)}(t,\vec{x}')  \right].
\end{equation}
After some integration by parts and using the identity
\begin{equation}
\grad \frac{1}{|\vec{x}'-\vec{x}|} = -\frac{1}{|\vec{x}'-\vec{x}|^3} (\vec{x}' - \vec{x}),
\end{equation}
we find 
\begin{eqnarray}
 \varphi^{(mix)}(t,\vec{x})
&=&
\frac{\alpha }{4(3+ 2\omega)\phi_0^{(out)}} \left( \frac{G_C}{2+\omega} \right)^2   
\nonumber 
\\
&& \times 
 \int d^3x'  \int d^3x'' 
 \frac{\rho(t,\vec{x}')}{|\vec{x}-\vec{x}'|^3}\frac{\rho(t,\vec{x}'')}{|\vec{x}' - \vec{x}''|^3} (\vec{x} - \vec{x}') \cdot(\vec{x'} - \vec{x}'').
\end{eqnarray}
To proceed we also need the term $|\grad \varphi^{(2,0)}|^2$ which is given by
\begin{eqnarray}
  \frac{  \alpha }{8(3+ 2\omega)\phi_0^{(out)}}   |\grad \varphi^{(2,0)}|^2  
&=&
  \frac{   \alpha }{8(3+ 2\omega)\phi_0^{(out)}}  \left( \frac{G_C}{2+\omega} \right)^2  
\nonumber 
\\
&& \times 
 \int d^3x'  \int d^3x'' \rho(x')\rho(x'') 
  \frac{(\vec{x} - \vec{x}')\cdot(\vec{x} - \vec{x}'')}{|x-x'|^3|x-x''|^3},
\end{eqnarray}
so that the full solution is
\begin{eqnarray}
\varphi^{(2,-2)}(t,\vec{x})  &=&
-  \frac{   \alpha  \left( \frac{G_C}{2+\omega} \right)^2 }{8(3+ 2\omega)\phi_0^{(out)}}  \int d^3x'  \int d^3x'' \rho(t,\vec{x}')\rho(t,\vec{x}'') 
\nonumber 
\\
&& \times 
\bigg\{ \frac{(\vec{x} - \vec{x}') \cdot(\vec{x} - \vec{x}'')}{|\vec{x}-\vec{x}'|^3|\vec{x} - \vec{x}''|^3} 
 - 2 \frac{(\vec{x} - \vec{x}') \cdot(\vec{x'} - \vec{x}'')}{|\vec{x}-\vec{x}'|^3|\vec{x}' - \vec{x}''|^3} 
\bigg\}.
\label{phi_22}
\end{eqnarray}

\subsubsection{The PPNV metric}
The new potential (\ref{phi_22}) that we have just found, is an example of what we call a Post-Newtonian-Vainshteinian potential.
To follow the spirit of the PPN formalism, let us define it slightly differently, i.e. without the constants that appear in front of the solution. We define the PPNV potential as
\begin{equation}
U^{(out)}_{V}  = 
 \int d^3x'  \int d^3x'' \rho(t,\vec{x}')\rho(t,\vec{x}'') \bigg\{
  \frac{(\vec{x} - \vec{x}') \cdot(\vec{x} - \vec{x}'')}{|\vec{x}-\vec{x}'|^3|\vec{x} - \vec{x}''|^3} 
 - 2 \frac{(\vec{x} - \vec{x}') \cdot(\vec{x'} - \vec{x}'')}{|\vec{x}-\vec{x}'|^3|\vec{x}' - \vec{x}''|^3} 
\bigg\},
\label{U_V}
\end{equation}
so that up-to $\O(2,-2)$ the solution for $\varphi^{(2)}$ is 
\begin{equation}
\varphi^{(2, \leq |-2|)} =  \frac{2\tilde{G}}{3+2\omega}U 
 -  \frac{   \alpha  \left( \frac{G_C}{2+\omega} \right)^2 }{8(3+ 2\omega)\phi_0^{(out)}}  U^{(out)}_{V} ,
\end{equation}
where the notation $\leq|-2n|$ means to sum all orders less  than or equal to $|-2n|$, i.e. $\varphi^{2, \leq |-2|}  = \varphi^{2,0}  + \varphi^{2,-2}$.

We may now use   (\ref{eq_h_00_2_standard}) and (\ref{eq_h_ij_standard})  to determine the metric up-to $\O(2,-2)$.
From  (\ref{eq_h_00_2_standard}) the metric solution to $\O(2,-2)$ is
\begin{equation}
 h_{00}^{(2, \leq|-2|)} =  2G_C U  +  2 g_V   G_C^3   U_{V}^{(out)}.
\label{eq_h_00_2_standard_V}
\end{equation}
The parameter $g_V$ is a PPNV parameter that for the Cubic Galileon is  
\begin{equation}
g_V = - \frac{\pi}{4} \left[\frac{M_p}{(2+\omega)\Lambda}\right]^3,
\end{equation}
  and for Brans-Dicke (GR) $g_V=0$.
Likewise from  (\ref{eq_h_ij_standard}) the metric solution to $\O(2,-2)$ is
\begin{equation}
h_{ij} =   \left[ 2 \gamma   G_C U   + \gamma_V G_C^3  U_V^{(out)} \right] \gamma_{ij},
\label{eq_h_ij_2_standard_V}
\end{equation}
where $\gamma_V$ is a new PPNV parameter that measures Vainshteinian corrections. In the Cubic Galileon theory we have
\begin{equation}
\gamma_V = -g_V =  \frac{\pi}{4} \left[\frac{M_p}{(2+\omega)\Lambda}\right]^3,
\end{equation}
and in Brans-Dicke or in GR $\gamma_V = 0$.
Note that as $\omega\rightarrow \infty$ we recover GR and the Vainshtein corrections vanish. This makes sense since in that limit $c_0 \rightarrow \infty$ and the scalar completely decouples. Indeed, by canonically normalising we see that the Galileon interaction is suppressed by a divergent strong coupling scale $
\Lambda_{strong}\sim \Lambda c_0^\frac32$.

\subsection{Dualizing the Cubic Galileon}
When the observer is placed close to the source the field self-interactions become strong and the standard approach used in the outside region breaks down. 
In order to proceed we must recast the original theory into a {\it classical dual} formulation ~\cite{GabadadzeHinterbichlerPirtskhalava2012,PadillaSaffin2012}. 
The basic idea is to introduce auxiliary variables 
``dual'' to the interaction terms in the standard action by a Legendre transform. It turns out that the equations of motion for the new set of variables
can be treated perturbatively since the expansion parameter $1/\Lambda$ is flipped to $\Lambda$. 

A technique to compute such a dual action was initially given in ~\cite{GabadadzeHinterbichlerPirtskhalava2012}, 
but in their dual variables the resulting 
equations involve non-analytic functional forms that make the mathematics quite cumbersome. A neater and more direct approach was put forth 
in ~\cite{PadillaSaffin2012}, where instead of using Legendre transforms as dual variables, the interaction terms are directly related to the
auxiliary variables by using Lagrange multipliers. In these variables simpler analytic equations
arise keeping all the advantages of the original approach. The dual action for the Cubic Galileon is given by 
\begin{equation}\label{actionDCG}
S_{\mbox{dual}} =  \frac{1}{16\pi G}\int d^4 x \sqrt{-g} \left\{\phi R - \frac{\omega}{\phi}(\nabla\phi)^2 
+\frac{1}{8\phi^3}\left(A^2\square\phi + 2Z A^\mu\nabla_\mu\phi \right)
- \frac{1}{4\sqrt{\alpha}} \frac{1}{\phi^3}  Z A^2
\right\},
\end{equation}
where $A_\mu$ is the dual field corresponding to $\nabla_\mu\phi$ and $Z$ is the dual field for $\square{\phi}$. 
By  extremizing the action with respect to the field and after some algebraic  manipulation we find the two relations between the dual fields and gradients of $\phi$
\begin{eqnarray}
 \nabla_\mu \phi &=& \alpha^{-\frac{1}{2}}  A_\mu,
\label{dualphi-A}
\\
 \square \phi  &=& \alpha^{-\frac{1}{2}} Z,
 \label{dualphi-Z}
\end{eqnarray}
the generalized Einstein equations 
\begin{eqnarray}
 \phi R_{\mu\nu} 
 &=& 8\pi G \left(  T_{\mu\nu} - \frac{1}{2} T g_{\mu\nu} \right) 
+  \alpha^{-\frac{1}{2}}  \bigg\{  \nabla_\mu A_\nu  + \frac{1}{2}  Z  g_{\mu\nu} +  \frac{\omega}{\phi} \alpha^{-\frac{1}{2}}   A_\mu A_\nu
\nonumber 
\\
&&
+  \frac{1}{8\phi^3} \left[ -   Z A_\mu A_\nu 
+   A_{(\mu}  \nabla_{\nu)} A^2
+ \frac{1}{2}    Z A^2  g_{\mu\nu}
-    3 \alpha^{-\frac{1}{2}}  \frac{A^2}{\phi}  A_\mu  A_\nu
\right]
\bigg\},
\label{EinstEqDual}
\end{eqnarray}
and the field equation for the dual fields (which is equivalent to the scalar equation in the standard formulation)
\begin{equation}
 - \square A^2 +2\nabla_\lambda(ZA^\lambda) =  - 64\pi G\phi^2T +  \alpha^{-\frac{1}{2}}
\left[  8(2\omega+3)\phi^2Z -  \frac{5}{\phi}A^\mu\nabla_\mu A^2
   +  \frac{1}{\phi}Z A^2+   \frac{9\alpha^{-\frac{1}{2}}}{\phi^2}A^4
\right].
 \label{dualAZ-T}
\end{equation}
Once again, we have manipulated the field equations to put them into a form more useful for performing the PPN expansion.
It can easily be checked that eliminating the dual fields from (\ref{EinstEqDual}) and (\ref{dualAZ-T}) using (\ref{dualphi-A}) and (\ref{dualphi-Z}) 
recovers the field equations in the standard formulation.

In the dual formulation the expansion parameter is $\alpha^{-1/2}$ which in this case will have order $\O(1,1)$.

\subsection{The PPNV expansion inside the Vainshtein radius}
As with the outside region, we expand the scalar field as
\begin{equation}
\phi = \phi_0^{(in)} \left( 1 + \varphi\right) = \phi_0^{(in)} \left( 1 + \varphi^{(2)} + \varphi^{(4)} + \ldots \right),
\end{equation}
where $\phi_0^{(in)}$ is a constant and $\varphi$ a perturbation. Note that due to the non-perturbative nature of the theory at the Vainshtein surface, 
the constant  $\phi_0^{(in)}$ will in general
not be equal to  $\phi_0^{(out)}$. However, they can in principle be related by appropriately matching the scalar $\phi$ across the two regions. 

Using the definitions of the dual fields (\ref{dualphi-A}) and (\ref{dualphi-Z})  we find that to PPN order $2$ 
\begin{eqnarray}
\phi_0^{(in)}\grad_i \varphi^{(2)} &=& \alpha^{-\frac{1}{2}} A_i^{(1)} = \phi_0^{(in)} \alpha^{-\frac{1}{2}} \grad_i B^{(1)}, 
\label{phiB2} 
\\
\phi_0^{(in)}\grad^2 \varphi^{(2)}  &=& \alpha^{-\frac{1}{2}} Z^{(1)}=   \phi_0^{(in)} \alpha^{-\frac{1}{2}} \grad^2 B^{(1)},
\label{phiZ2} 
\end{eqnarray}
where we have defined the scalar field $B$ whose gradient gives the dual field $A_\mu$.
Thus, in general, the orders of $\varphi$ will be one greater than the orders of the $B$ field (either PPN or Vainshtein order).
Notice that as the constant $\alpha^{-\frac{1}{2}}$ is of order $\O(1,1)$, it increases both the PPN order and the Vainshtein order 
of any terms multiplying it by one. This means that the lowest PPN order for the scalar perturbation $\varphi$ is $2$ 
while the lowest Vainshtein order for the scalar field perturbation $\varphi$ is $1$ (given that the lowest possible Vainshtein order for the $B$ field is zero).

We now consider the generalized Einstein equations to $\O(2)$ in PPN.

\subsubsection{$h_{00}$ to $\O(2)$}

Taking the $R_{00}$ equation (\ref{EinstEqDual}) and expandng to $\O(2)$ using also (\ref{R_00_full}) we find
\begin{eqnarray}
\frac{1}{2}  \grad^2  h_{00}^{(2)}  &=& -4\pi G_N \rho   + \frac{1}{2}\alpha^{-\frac{1}{2}} \grad^2 B^{(1)},
\end{eqnarray}
where we have defined
\begin{equation}
 G_N = \frac{G}{\phi_0^{(in)}}.
\end{equation}
We can determine $ h_{00}^{(2)}$ completely in terms of the new potential $B^{(1)}$. The answer is
\begin{eqnarray}
 h_{00}^{(2)}  = 2G_N U   + \alpha^{-\frac{1}{2}} B^{(1)}.
\label{eq_h_00_dual}
\end{eqnarray}

Since the lowest Vainshtein order for  $ B^{(1)} $ is $0$ taking the limit $\alpha^{-\frac{1}{2}}\rightarrow 0$ recovers General Relativity.
Thus the Vainshtein mechanism is directly manifested in the dual formalism. 
In other words, to $\O(2,0)$ the $B^{(1)}$ term does not contribute and we recover the usual Newtonian potential
\begin{equation}
 h_{00}^{(2,0)}  =  2 G_N U.
\end{equation}
Hence $G_N$ has the meaning of the observable Newtonian gravitational constant inside the Vainshtein radius, i.e. the Newtonian constant measured by a table-top experiment.

Further out away from the source, the Vainshteinian potential $ B^{(1)}$ will start to have an effect and we need to take into account the Vainshteinian 
corrections.  This is achieved via the scalar equation to $\O(2)$ (see Eq. \ref{eq_B_1})

\subsubsection{$h_{ij}$ to $\O(2)$}
Taking the $R_{ij}$ equation  (\ref{EinstEqDual}) and expandng to $\O(2)$ using also  (\ref{R_ij_full}) we find
\begin{eqnarray}
&&
\frac{1}{2} \grad_k   \left[ \grad_i h^k_{\;\;j} + \grad_j h^k_{\;\;i}\right]
 - \frac{1}{2}  \grad^2 h_{ij}
 + \frac{1}{2} \grad_i \grad_j (  h_{00}^{(2)} - h)
 = 4\pi G_N  \rho \gamma_{ij} 
\nonumber 
\\
&&
\qquad 
\qquad 
 + \alpha^{-\frac{1}{2}} \left[ \grad_i \grad_j B^{(1)}   + \frac{1}{2} \grad^2 B^{(1)}   \gamma_{ij} \right].
\label{eq_R_ij_dual}
\end{eqnarray}
Imposing the gauge fixing
\begin{equation}
 \grad_k h^k_{\;\;i} =  \grad_i \left( \frac{1}{2} h -  \frac{1}{2} h_{00}^{(2)}  + \alpha^{-\frac{1}{2}} B^{(1)}\right),
\label{eq_GF_dual_1}
\end{equation}
turns (\ref{eq_R_ij_dual}) into
\begin{eqnarray}
 - \frac{1}{2}  \grad^2 h_{ij}
 &=& 4\pi G_N  \rho \gamma_{ij}  +  \frac{1}{2}  \alpha^{-\frac{1}{2}} \grad^2 B^{(1)}   \gamma_{ij},
\label{eq_R_ij_dual_GF}
\end{eqnarray}
so that we obtain $h_{ij}$ as
\begin{eqnarray}
  h_{ij} &=&   \left( 2G_N U   -  \alpha^{-\frac{1}{2}}  B^{(1)} \right)   \gamma_{ij},
\label{eq_h_ij_dual}
\end{eqnarray}
We notice that no new potential (apart from $B^{(1)}$) arises. Once again as $\alpha^{-1/2} \rightarrow 0$ we recover GR, i.e.
\begin{equation}
  h_{ij}^{(2,0)} =    2G_N U     \gamma_{ij}.
\end{equation}

\subsubsection{Scalar equation to $\O(2)$} 
Just as in the case outside the Vainshtein radius, in order to proceed further we need the scalar equation to $\O(2)$. We find
\begin{equation} 
 \grad^2 \left(|\grad B^{(1)}|^2    \right) -   2 \grad \cdot\left(  \grad^2 B^{(1)}   \grad B^{(1)} \right)  = -64\pi G    \rho - 8\alpha^{-1/2}   (2\omega + 3) \phi_0^{(in)}  \grad^2 B^{(1)},
\label{eq_B_1} 
\end{equation}
we then expand the dual field in Vainshtein orders, i.e.
\begin{equation}
 B^{(1)} = B^{(1,0)} +  B^{(1,1)} +  B^{(1,2)} + \ldots  =  \sum_{n=0}^{\infty} B^{(1,n)}.
\end{equation}
Taking the lowest order we get an equation for the dual field $B^{(1,0)}$
which is 
\begin{equation}
 \grad^2 \left( |\grad B^{(1,0)}|^2     - 16 G U\right) -   2 \grad \cdot\left(  \grad^2 B^{(1,0)}   \grad B^{(1,0)} \right)  =  0. 
 \label{eq_B_10}
\end{equation}
However, at the same limit,  
the terms involving $B^{(1)}$ disappear in the metric solutions (\ref{eq_h_00_dual}) and (\ref{eq_h_ij_dual}). What happens is that the solution for the metric to Vainshtein order $0$
is simply General Relativity while the field $B^{(1)}$ decouples and obeys its own differential equation with  no bearing on the metric or matter evolution.

But the field $B^{(1)}$  will eventually have a bearing on the metric.  Solving (\ref{eq_B_10}) and inserting into (\ref{eq_h_00_dual}) and (\ref{eq_h_ij_dual}) gives 
us the  first correction to the GR metric, i.e. the metric solution  up-to $\O(2,1)$.
Unfortunately a complete analytic solution to (\ref{eq_B_10}) is (to our knowledge) impossible, except in idealized situations like spherical symmetry. 
Indeed, by simple shuffling of the terms involved, (\ref{eq_B_10}) can be re-written in the form
\begin{eqnarray}
  \grad^i \grad^j B^{(1,0)}  \grad_i \grad_j B^{(1,0)} -   \left(  \grad^2 B^{(1,0)} \right)^2 &=&  -32\pi G \rho.
\label{Monge_Ampere}
\end{eqnarray}
We recognise the above equation as a Monge-Ampere like equation whose general solution is unknown, and one has to resort to numerics.
Thus, it is impossible to find this new Vainshteinian potential as a closed form integral as we have done in the case outside the Vainshtein radius.

\subsubsection{Reconstructing the metric: the $\O(2)$ metric solution to all Vainshtein orders}

Fortunately, the polynomial structure of (\ref{eq_B_1}) lends itself to the use of perturbation theory. Once the lowest order solution is found,
for instance $B^{(1,0)}$ then higher orders $B^{(1,n)}$ will obey a linearized  equation coming from  (\ref{eq_B_1}). More specifically,  
expanding order-by-order (\ref{eq_B_1})  may be re-written as
\begin{eqnarray}
\mathcal{L}(B^{(1,0)},B^{(1,0)}) &=& 64 \pi G\rho,
\nonumber
\\
\mathcal{L}(B^{(1,1)},B^{(1,0)})+\mathcal{L}(B^{(1,0)},B^{(1,1)})&=&  8\alpha^{-\frac{1}{2}}(2\omega+3) \phi_0^{(in)}  \nabla^2B^{(1,0)}.
\label{B_lin}
\\
\ldots
\nonumber
\end{eqnarray}
The operator $\mathcal{L}$ is the bi-linear operator $\mathcal{L}: C^\infty \times C^\infty \rightarrow C^\infty$ defined by
\begin{equation}
\mathcal{L}(u,v) = 2\grad\cdot \left[(\grad^2 v)  \grad u  \right]  - \grad^2 (\grad u \cdot \grad v),
\end{equation}
for two arbitrary functions $u,v\in C^{\infty}$. The operator  $\mathcal{L}$ is a
 non-symmetric ($\mathcal{L}(u,v) \neq \mathcal{L}(v,u)$) bi-linear form, i.e. it obeys the
 properties $  \mathcal{L}(u+v,w) =  \mathcal{L}(u,w) + \mathcal{L}(v,w)$ and 
$\mathcal{L}(\lambda u,v) = \mathcal{L}(u,\lambda v) $.
Then we can formally write the solutions to the hierarchy of linearizations  (\ref{B_lin}) by considering the operator
$\hat{\mathcal{L}}: C^\infty\rightarrow  C^\infty$ defined by
\begin{equation}
\hat{\mathcal{L}}\mathcal{L}(u,v) = uv .
\label{hatL}
\end{equation}
The operator $\hat{\mathcal{L}}$ is distributive with respect to addition and to multiplication by a constant that ensures 
that $\mathcal{L}$ is invertible and a solution can be constructed.

In the spirit of the PPN potentials and also the Vainshteinian potential $U^{(out)}_V$ 
the last considaration is to rescale the potentials $B^{(1,n)}$ to factor out any model-dependent parameters. We therefore define the series of
Vainshteinian potentials  for each Vainshtein order $n$ as
\begin{equation}
 U^{(in)}_{n}  =  \frac{1}{4\sqrt{G}} \left[  \frac{\sqrt{G\alpha}}{ 2 (2\omega+3) \phi_0^{(in)}}  \right]^n  B^{(1,n)},
\end{equation}
so that the  hierarchy of linearizations (\ref{B_lin}) becomes
\begin{eqnarray}
\mathcal{L}(U^{(in)}_{0},U^{(in)}_{0}) &=& 4 \pi \rho,
\nonumber
\\
  \mathcal{L}( U^{(in)}_{1},U^{(in)}_{0} )+\mathcal{L}(U^{(in)}_{0} , U^{(in)}_{1})&=&   \nabla^2U^{(in)}_{0},
\label{U_lin}
\\
\ldots
\nonumber
\end{eqnarray}
The above hierarchy can then in principle be solved order-by-order where now the only input is the matter density $\rho$.
Formally this is achieved via the operator $\hat{\mathcal{L}}$ that ensures that the solutions exist:
\begin{eqnarray}
U^{(in)}_{0} &=& \sqrt{4\pi  \hat{\mathcal{L}}( \rho ) },
\nonumber
\\
 U^{(in)}_{1}&=& \frac{1}{2 U^{(in)}_{0} } \hat{\mathcal{L}} (  \nabla^2U^{(in)}_{0} ),
\label{U_lin_sol}
\\
\ldots
\nonumber
\end{eqnarray}

 Once this is done the metric can then be reconstructed as
\begin{eqnarray}
 h_{00}^{(2)}  &=&  2G_N U   + 2 \sum_n    g^{(in)}_n  G_N^{-n} U^{(in)}_n,
\\
  h_{ij} &=&   \left( 2G_N\gamma U    + 2 \sum_n     \gamma^{(in)}_n  G_N^{-n}   U^{(in)}_n   \right)   \gamma_{ij} ,
\end{eqnarray}
where we have introduced the PPNV parameters $g^{(in)}_n$ and $\gamma^{(in)}_n $ along with the PPN parameter $\gamma$. For GR we have $\gamma=1$ and $g^{(in)}_n  =  \gamma^{(in)}_n =0$,
for the Brans-Dicke theory  $\gamma= \frac{1+\omega}{2+\omega} $ and $g^{(in)}_n  =  \gamma^{(in)}_n =0$ while for the Cubic Galileon we have $\gamma=1$ and
\begin{eqnarray}
g^{(in)}_n  =  -  \gamma^{(in)}_n  =  \frac{2^{1+n} (2\omega+3)^{n} }{(8\pi)^{\frac{n+1}{2}}}  \left[\frac{\Lambda\sqrt{\phi_0^{(in)}}}{M_p^{(nom)}}\right]^{\frac{3(n+1)}{2}},
\end{eqnarray}
where $M_p^{(nom)} = \frac{1}{\sqrt{8\pi G_N}}$ is the nomimal Planck mass (defined using the measured Newton's constant $G_N$ rather than $G$.

\section{Back to spherical symmetry}
 We have determined the PPNV expansion for the Cubic Galileon theory to $\O(2)$ in PPN both inside and outside the Vainshtein radius in terms of general potentials. Let us now
compare the expansion with the results for spherical symmetry that we have found in section \ref{sec_spherical}. We start from the expansion outside the Vainshtein radius.
The second term inside the integral of the Vainshteinian potential (\ref{U_V}) is antisymmetric in the exchange of $\vec{x}'$ and $\vec{x}''$ which means that it will vanish in any spherically symmetric situation. Using $\rho(t,\vec{x}) = M \delta^{(3)}(\vec{r}) =\frac{M}{4\pi} \frac{\delta(r)}{r^2}$ we evaluate the integral as $U_V = \frac{M^2}{r^4}$. Since
the Newtonian potential is $U = \frac{M}{r}$, we determine the metric solution as 
\begin{eqnarray}
  h_{00} &=& \frac{2G_C M}{r}  \left[1 + \frac{g_V G_C^2 M}{r^3}\right] = \frac{2G_C M}{r}  \left[1 - \beta_{(out)} \left(\frac{r_V}{r}\right)^3\right],
\\
  h_{ij} &=& \frac{2G_C M}{r}  \left[ \frac{1+\omega}{2+\omega} + \frac{\gamma_V G_C^2 M}{r^3}\right] \gamma_{ij}= \frac{2G_C M}{r}  \left[ \frac{1+\omega}{2+\omega} + \beta_{(out)} \left(\frac{r_V}{r}\right)^3\right]\gamma_{ij},
\end{eqnarray}
where $\beta_{(out)} =  \frac{1}{64\pi (3+2\omega)^2(2+\omega) (\phi^{(out)}_0)^2}$ which is the spherically symmetric solution in  section \ref{sec_spherical} (only now we have also
found the metric component $h_{ij}$. 

We turn now to the inner region. The operator $\mathcal{L}$ for spherical symmetry gives
\begin{equation}
\mathcal{L}(u,v) = u' v''' -   u''' v' +\frac{ 2}{ r} (v' u''  + 3 u'v'') +  \frac{4}{r^2} u' v' .
\end{equation}
For the Vainshtein order $0$ case, $u = v = U^{(in)}_0$ so that $\mathcal{L}(u,u) = \frac{4}{r^2} \frac{d}{dr} [r (u')^2   ] = \frac{ M}{r^2} \delta(r)$ which gives the solution
$ U^{(in)}_0 = \pm\sqrt{Mr} $.  Choosing the minus sign of the solution, i.e. $U^{(in)}_0 = -\sqrt{M r}$, leads to the metric solution
\begin{eqnarray}
 h_{00}^{(2)}  &=&  2G_N \frac{M}{r} \left[1 -  4 \sqrt{2\pi} (\phi_0^{(in)})^{3/4} \left(\frac{r}{r^{(nom)}_V}\right)^{3/2} \right],
\\
 h_{ij}  &=&  2G_N \frac{M}{r} \left[1 +  4 \sqrt{2\pi}   (\phi_0^{(in)})^{3/4} \left(\frac{r}{r^{(nom)}_V}\right)^{3/2} \right] \gamma_{ij},
\end{eqnarray}
where we have defined the nomimal Vainshtein  radius $r_V^{(nom)} =  \frac{1}{\Lambda} \left(\frac{M}{M_p^{(nom)}}\right)^{1/3}$. 

The plus/minus signs in the solution for $U^{(in)}_0$ has to do with the existence of solution branches in the field equations. This is the same issue
that we encountered in (\ref{chi_p_spher_sym}) where at that instance we chose the plus sign in front of the square root.
 Had we chosen the opposite sign, we would have recovered
the solution found here with the plus sign, i.e. $U^{(in)}_0 = \sqrt{M r}$ for the inner region $r<r_V$. However, for the outer region, the leading order term would have then been $\chi \sim r^2$, rather than $\sim \frac{1}{r}$, with the constant of proportionality depending only on $\alpha$
 and not the mass of the source. 
Therefore this branch of the solution is not present in our PPNV expansion in the outer region $r>r_V$ for the reason that it is not sourced by matter 
(and therefore cannot be recovered from (\ref{U_V})) but also due to our assumption of a background Minkowski metric, an assumption which would be  violated by this solution branch. Perhaps, one has to expand around de Sitter space in order to have this branch in the solution space.

Let us now turn to the values $\phi_0^{(in)}$ and $\phi_0^{(out)}$. In principle, one can rescale units, for instance, $\Lambda \rightarrow \Lambda \sqrt{\phi_0^{(out)}}$
and similarly for all other mass (and length) scales. Then one can set either of  $\phi_0^{(in)}$ or  $\phi_0^{(out)}$ (but not both) 
to unity by simple unit re-definition. It is a matter of choice which one, but for what follows we choose to set $\phi_0^{(in)} = 1$ as it is the
most relevant to the solar system (in which case $G_N = G$ and $M_p^{(nom)} = M_p$).

In analogy with the standard PPN formalism, we may try to check the effect of the parameters $g_n$ and $\gamma_n$ on null geodesics. Using the geodesic equation for light we find 
 that the Shapiro time-delay and the light deflection angle are both proportional to the combinations $g_n + \gamma_n$ for all orders $n$.
 While these may be used to place constraints on the combination $g_n + \gamma_n$, clearly, 
 the Cubic Galileon terms can not be constrained using light, although a disformally coupled Galileon would presumably give a different result. Of course,
by tracking material bodies, it may also be possible to place constraints on the PPNV parameters. We leave these considerations for a future work.

For completeness, let us give a rough estimate of the deviations from GR in the solar system. In figure 1 we display the functions $h_{00}^{(2)}$ and $h_{ij}$ versus $r$
as well as the residuals from GR. We assume a spherically symmetric solar system with all matter concentrated in the sun. Then $M_\odot/M_p = 1.12 \times 10^{57}$
so that $r_V = 1.02 \times 10^{-5} \, \frac{1 eV}{\Lambda} \; $au.
 For a choice of  $\Lambda \sim (M_{pl}H_0^2)^\frac13 \sim 1/1000\text{km} \sim  2\times 10^{-13}eV$ ($r_V \sim 5 \times 10^7 $au) (in order  that the scalar modifies the late time cosmology from the present epoch onwards), 
 this theory predicts deviations of the order 
 $\sim 10^{-10}$ at the orbit of Jupiter and $\sim 10^{-9}$ at the orbit of Neptune. It is interesting also that the Vainshtein radius of the Sun for this choice of $\Lambda$ 
is larger than the distance to the nearest star.

\begin{figure}[h]
\center
\epsfig{file=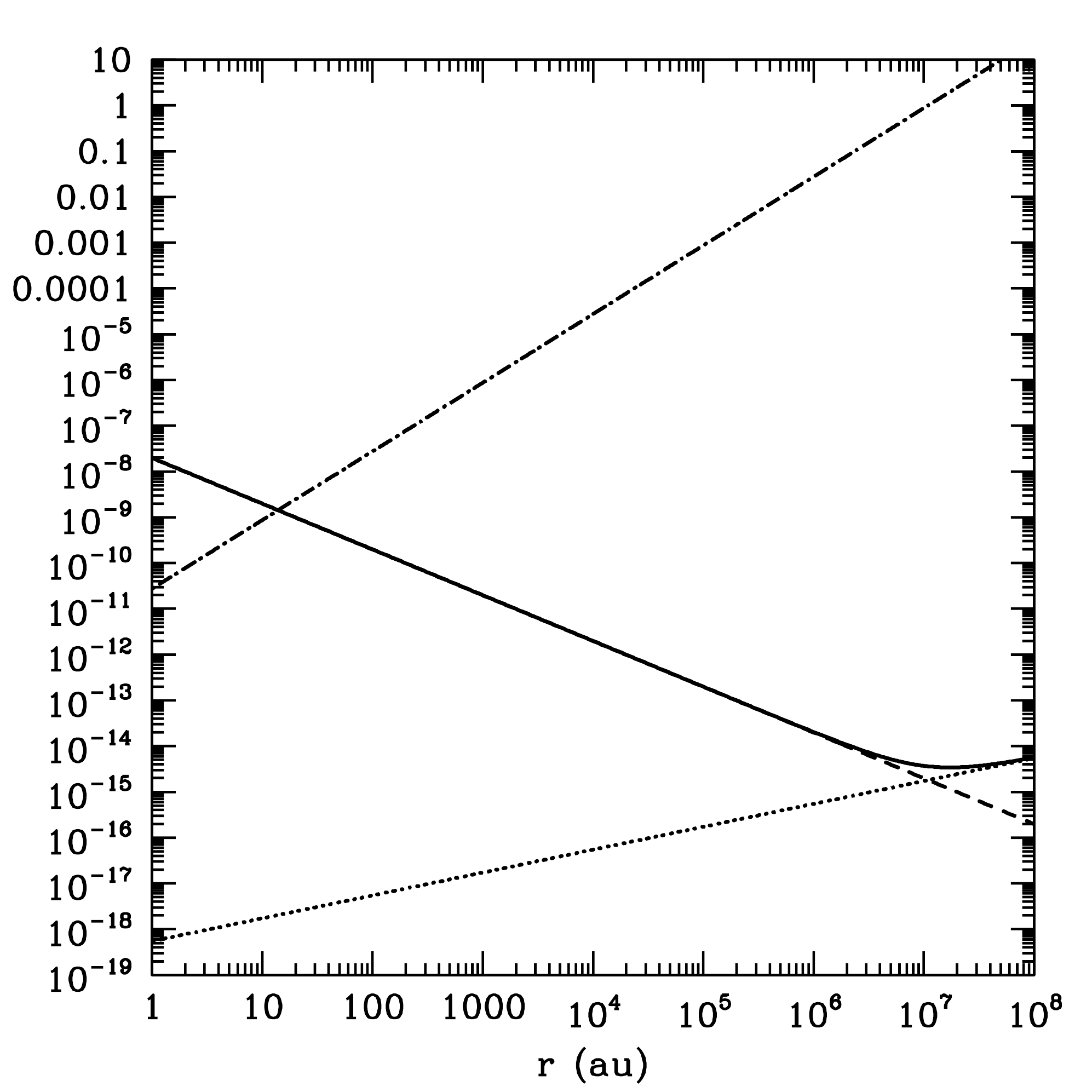, width= 75 mm, height= 70 mm}
\caption{Solutions for the potential $h_{ij}$ for the Cubic Galileon inside the Vainshtein radius. 
We display the Newtonian term (dash), the Vainshteinian correction (dotted), the total potential (solid) and $h_{ij}^{(2,1)}/h_{ij}^{(2,0)}$ (dash-dot).}
\end{figure}

\section{Conclusion}
We have presented a general scheme for calculating the metric for theories exhibiting Vainshtein screening, in which there is a unique scale beyond the Schwarzschild radius where non-linearities begin to become important. This scheme, the Parametrized Post-Newtonian Vainshteinian expansion, is an expansion in two small parameters, namely, small velocities and a parameter that determines the strength of the Vainshteinian corrections. We saw that
such theories have two regimes: inside the Vainshtein radius and outside. One has to perform a separate expansion in each regime. In particular,
the inner regime requires the use of a dual and classically equivalent action that inverts the scale $\Lambda$ and makes the expansion in a small parameter feasible.

The long term goal of the  PPNV framework is to help determine the compatibility of the relevant theories with solar system and other strong-field data and we expect it to play
a fundamental role in constraining them independently from cosmology.  This paper should be regarded as the first in developing a full formalism. Whilst we have explained the formalism for the class of theories under consideration i.e. those with a single Vainshtein radius, we have only presented this explicitly for the Cubic Galileon, working up to PPN order 2. 
 We expect new  PPNV potentials to  arise as we introduce more operators and/or increase the order of our expansion, just as is the case with PPN. In a separate calculation we will present the expansion  to
$\O(4)$ in PPN for the Cubic Galileon. It would also be pertinent to identify PPNV potentials for the higher Galileons, leading order  k-essence terms, and for massive gravity (to name a few). 

For the Cubic Galileon case, we have explicitly demonstrated the working of the Vainshtein mechanism. Interestingly,  we found that the theory tends to Brans-Dicke far outside the Vainshtein radius, i.e. in the cosmological regime, while it tends to GR deep inside the Vainshtein radius, where solar-system tests lie. Thus, the Cubic Galileon is a concrete example where solar-system constraints on Brans-Dicke theory do not apply but rather one has to use cosmological constraints~\cite{AvilezSkordis2013}. 

A more challenging extension would be to allow for multiple macroscopic scales, multiple Vainshtein  radii if you like. This will obviously mean expanding in more than two variables. Note that the classical dual can be formulated  to cope with mutliple strong coupling scales (which would in turn lead to multiple Vainshtein radii). Difficulties could occur, however, when interactions suppressed by the same strong coupling scale kick in at different macroscopic scales due to their differing derivative structure. It is not entirely clear how the classical dual will capture this as an expansion in the appropriate Vainshtein radii. Presumably we will have to extend the classical dual itself in order to cope with this.

Finally, let us comment on the so called {\it self-screening effect} which has been argued to play an important role for Vainshtein systems \cite{selfsc1, selfsc2}, and suggests, for example,  that the moon cannot be treated as a test body in the field profile of the earth.  We essentially agree with this. The point is that our method allows us to extract a perturbative solution for the scalar profile and the geometry inside and outside the Vainshtein radius for the entire system. If we take our probe to be part of the source in that calculation, then self-screening effects should be accommodated although to be practical one may wish to include an additional expansion is terms of, say, the relative mass scales in a two body problem. The non-linear nature of the inner PPNV expansion would be key to incorporating self-screening. In contrast, if we do not include our probe as part of the source, then we are genuinely treating it as a test particle and ignoring any self-screening.
 
To be more explicit, a small body orbiting the Sun can indeed be taken as part of the source so that $\rho  = \rho_\text{sun} + \rho_\text{body}$.  We then solve our field equations and calculate the corrections to the potential as explained in this paper. Of course there is no spherical symmetry but that is a calculational problem rather than an issue with our formalism. Once we have the metric,  the small body will certainly follow the geodesics of that metric just like a test body. The key point, however, is that it will not follow the geodesics of a metric calculated from $\rho_\text{sun}$ alone.  This is the correct statement regarding both the self-screening papers and our work. How big the effect is, and how one can re-cast it as a violation of the SEP, are two different questions, but two answerable questions within our formalism.

\acknowledgments 

We thank Tessa Baker, Javier Chagoya, Pedro Ferreira, Adam Moss and Gustavo Niz for useful discussions and comments.
 A.A. acknowledges support from CONACYT.  A.P acknowledges support from the Royal Society, and P.M.S. acknowledges support from STFC.
C.S. acknowledges initial support from the Royal Society and further support from the European Research Council.
The research leading to these results has received funding from the European Research Council 
under the European Union's Seventh Framework Programme (FP7/2007-2013) / ERC Grant Agreement n. 617656 ``Theories
 and Models of the Dark Sector: Dark Matter, Dark Energy and Gravity''. 

\bibliographystyle{JHEP}
\bibliography{references}

\end{document}